\documentclass[12pt]{article}
\usepackage{graphicx}  

\newcommand{\beq}{\begin{equation}}
\newcommand{\eeq}{\end{equation}}
\newcommand{\bea}{\begin{eqnarray}}
\newcommand{\eea}{\end{eqnarray}}

\newcommand{\ave}[1]{\langle {#1} \rangle}

\newcommand{\xx}{\ave{x^2}} 
\newcommand{\pp}{\ave{p^2}}
\newcommand{\px}{\ave{p\,x}} 
\newcommand{\xxx}{\ave{x^3}} 
\newcommand{\ppx}{\ave{p^2\,x}} 
 
\newcommand{\xxxx}{\ave{x^4}} 
\newcommand{\pppp}{\ave{p^4}} 
\newcommand{\ppxx}{\ave{p^2\,x^2}} 
\newcommand{\nxxx}{\ave{:x^3:}} 
\newcommand{\nppx}{\ave{:p^2\,x:}} 
\newcommand{\nxxxx}{\ave{:x^4:}} 
\newcommand{\npppp}{\ave{:p^4:}} 
\newcommand{\nppxx}{\ave{:p^2\,x^2:}} 
\begin{document}  
  
\title{A consistent approximation scheme beyond RPA for bosons}  
\author{D. Davesne, M. Oertel, H. Hansen}  
\maketitle  

\begin{abstract}  
In this paper, we develop a consistent extension of RPA for bosonic systems. In
order to illustrate the method, we consider the case of the anharmonic
oscillator. We compare our results with those obtained in mean-field and
standard RPA approaches, with the exact ones and show that they are
very close to the exact ones. 
\end{abstract}  

~{IPN Lyon, 43 Bd du 11 Novembre 1918, F-69622 Villeurbanne Cedex}   

\begin{flushleft}  
LYCEN 2002-30  
\end{flushleft}

\section{Introduction}  
In many domains of physics perturbation theory fails to describe
physical phenomena correctly. For example, bound states or collective
excitations cannot be described perturbatively. In particular in the
low-energy regime of strong interactions non-perturbative effects play
an important role. Although much progress has been
made in understanding the hadronic world, it is still indispensable to
develop non-perturbative, symmetry conserving methods.

In the last few years much effort has been made to apply
non-perturbative techniques well established in many-body theory to
quantum field theory (see, e.g.,
Refs.~\cite{HCPT95,ACSW96,ASW97,HCDS02}). In this article we will
concentrate on approaches related to the random phase approximation
(RPA). In its standard form the RPA is known to respect
symmetries. For example, in Ref.~\cite{ACSW96} it has been
demonstrated within a linear $\sigma$-model that the RPA fluctuations
generate the Goldstone bosons related to the spontaneously broken
global (chiral) symmetry. At mean-field level, on the contrary, these
soft modes do not emerge. Thus the RPA seems to be a method very well
suited to treat non-perturbative systems with spontaneously broken
symmetries. Nevertheless it has some shortcomings: It is not of
variational character and therefore does not necessarily fulfill the
Rayleigh-Ritz criterion. In addition, it is not self-consistent, and
therefore the intermediate states (corresponding to the mean-field
quasiparticles) do not coincide with the final RPA particles. In
particular, even if the RPA states show up as massless Goldstone
modes, the intermediate states are massive. This has been pointed out
by Aouissat {\it et al.}~\cite{ASW97,A00} in the context of pionic
excitations in an $O(N)$-model.

It is obvious that this self-consistency problem can be circumvented in
$1/N$-expansions~\cite{ASW97,A00,OBW01}. On the other hand, a fully
self-consistent approximation such as Self-Consistent RPA
(SCRPA)~\cite{RDS95} is desirable. This method has
already been applied successfully to fermionic systems, e.g., to the Hubbard
model~\cite{SS99} and the seniority model~\cite{RSBCD01}. For a bosonic
field theory a first attempt in this direction has been made in
Ref.~\cite{HCDS02}. There some of the correlations are included, which are
not present in the mean-field ground state used in standard RPA. This
procedure, however, causes serious problems: In particular it does not
preserve covariance, an essential
ingredient of a relativistic quantum field theory.

In order to establish a consistent method and to clearly state the
problems related to the treatment of a bosonic system in an extended
RPA formalism we will reduce the dimension of the system. Instead
of a $\lambda \Phi^4$-theory as in Ref.~\cite{HCDS02} we
will study a one-dimensional anharmonic oscillator. Of course this is a
purely quantum mechanical, non-relativistic system and we can
certainly not tackle the problem of covariance. Nevertheless, we can
isolate some of the inherent contradictions related to
bosonic systems. It has to be examined within a future work whether
the method presented here is able to preserve covariance.

Our toy model, the anharmonic oscillator has the 
advantage, that exact results are known and that we can directly
compare the results obtained in different approximation schemes with
those obtained by solving the Schr\"odinger equation. It has already
been used as a test ground, e.g., for coupled cluster
techniques~\cite{BF88} or first attempts within SCRPA~\cite{DS91}. 

The article is organized as follows: We begin with presenting the RPA
equations in Sec.~\ref{RPAeqs}. In Sec.~\ref{RPAnormal}, we
review the Hartree-Fock (mean-field) approximation together with the
standard RPA. The renormalized RPA (r-RPA) and its problems are
discussed in Sec.~\ref{RRPA}. In Sec.~\ref{modifRRPA}, we present
our extended r-RPA approach. Sec.~\ref{numerics} is devoted to a
discussion of the numerical results, in particular for the ground
state energy, obtained within the different approximation
schemes. Finally, in Sec.~\ref{conclusion}, we will draw our
conclusions.

\section{The RPA equations}
\label{RPAeqs}

First of all, we have to state our starting point, the RPA equations.
There are mainly two different ways to derive the RPA equations: the
equation of motion method due to Rowe~\cite{R68} and the
Dyson-Schwinger-approach \cite{SE73}. To derive the former, we suppose
that we can generate an excited state $|\nu\rangle $ by the action of
an operator $Q^{\dagger}_{\nu}$ on the ground state
$|\mathrm{gs}\rangle $: $|\nu\rangle =
Q^{\dagger}_{\nu}|\mathrm{gs}\rangle $, $|\mathrm{gs}\rangle $ being
the vacuum of $Q_{\nu}$. Obviously $Q^{\dagger}_{\nu}$ is a highly
non-trivial many-body operator. Depending on the nature of
$|\nu\rangle $ (excited state with the same number of particles or
excited state with a different number of particles),
$Q^{\dagger}_{\nu}$ can be seen as a superposition of (non-hermitian)
one, two or more body operators. In general,
$Q^{\dagger}_{\nu}$ is expressed in the following form:
\beq Q^{\dagger}_{\nu}
= \sum _{\alpha} X^{\nu}_{\alpha} A^{\dagger}_{\alpha}~,  
\eeq 
where $A^{\dagger}_{\alpha}$ contains all the chosen excitation
operators expressed in terms of $a_i$ and $a^{\dagger}_i$ (annihilation
and creation operators) and $\alpha$ represents all quantum
numbers. Moreover, minimization of the energy $E_{\nu} = \langle
\nu|\mathrm{H}|\nu\rangle$ with respect to a variation $\delta Q_\nu$ leads to
a system of equations whose
solution gives the amplitudes $X^{\nu}_{\alpha}$
and the excitation energies:
\beq
\langle \mathrm{gs}|\left[\delta
Q_\nu,[\mathrm{H},Q^\dagger_\nu]\right]|\mathrm{gs}\rangle = (E_\nu- E_0)
\langle\mathrm{gs}| [\delta Q_\nu,Q^\dagger_\nu]|\mathrm{gs}\rangle~,
\eeq
where $E_0$ represents the energy of the ground state $|\mathrm{gs}\rangle$.

In the Dyson-Schwinger approach, which will be used in this paper, one
starts from the definition of a time-ordered Green's function at zero
temperature in equilibrium (the generalization to finite temperature
is direct):
\beq
G_{\alpha \beta}(t,t')=-i\langle \mathrm{gs}| T\left(A_{\alpha}(t)\, \,
A^{\dagger}_{\beta} (t')\right)|\mathrm{gs}\rangle~,
\eeq
where $T$ is the time
ordering operator and $A$ and $A^{\dagger}$ depend on time via $A(t) =
e^{iHt} A(0) e^{-iHt}$, where H is the exact Hamiltonian. In
equilibrium the above two-time Green's function\footnote{Note that we
deal with a two-time Green's function. That means that, if, e.g., $A =
a^\dagger a^\dagger$, both particles are created at the same time.}
is a function of the time difference $t-t'$ so that the Fourier
transform $G(E)$ depends only on one frequency. Our aim is to derive a Dyson
equation for $G(E)$. If we define
\begin{equation}
G_{\alpha \beta}(t,t')=\int\,{dE\over 2\pi}\,e^{-i\,E\,(t-t')}\,
G_{\alpha \beta}(E)~,
\end{equation}
the result is
\begin{equation}
E\,G_{\alpha \beta}(E)={\cal N}_{\alpha \beta}\,+\,
\sum_{\beta ',\gamma}{\cal  DC} _ {\alpha \beta '}{\cal N}^{-1}_{\beta '
\gamma} \,G_{\gamma\alpha}(E)~,
\label{RPAEQC}
\end{equation}
with
\bea
{\cal N}_{\alpha \beta} &=& \langle \mathrm{gs}| [A_{\alpha},
A^{\dagger}_{\beta}] |\mathrm{gs} \rangle~, \nonumber \\
{\cal DC}_{\alpha \beta}&=&\langle \mathrm{gs}| [[A_{\alpha}, \mathrm{H}],
A^{\dagger}_{\beta}] |\mathrm{gs} \rangle~. \qquad
\label{dc}
\eea
It has to be noted that the operators entering the
above Eq.~(\ref{dc}) have to be taken at equal times since the
truly dynamical contribution has been omitted (see, e.g., the
derivation in Refs.~\cite{SS99,SDR97}).

One can show that the two methods mentioned above are strictly equivalent
provided that~\cite{SDR97}
\beq
\langle \mathrm{gs}| [\mathrm{H},[A_{\alpha},A^{\dagger}_{\beta}]] |\mathrm{gs} \rangle = 0~.
\eeq
Of course, if $|\mathrm{gs}\rangle$ is the exact ground state, i.e., an
eigenstates of the Hamiltonian, this relation is automatically satisfied.
Within an approximation this needs, however, not to be the case.
Nevertheless, in most practical situations, including our treatment of the
anharmonic oscillator, it can be shown that both methods are indeed
equivalent. 

Obviously it is not possible to include the complete Hilbert space of
excitation operators in $A$ and it is necessary to restrict it. In
principle the approximation can straightforwardly be refined by
enlarging the space of excitation operators contained in $A$. But, in
practice this often turns out to be a very difficult task because, on
one hand, the dimension of the RPA matrix to be diagonalized increases
and, on the other hand, there exist consistency problems for the
theory itself (see the next section).

Another crucial point in solving the above RPA equations,
Eq.~(\ref{RPAEQC}), is the way in which the expectation values are
determined since the exact ground state is not known. It is however
clear that it strongly influences the quality of the approximation. In
standard RPA the exact ground state is usually replaced by a
mean-field (Hartree-Fock) one. But we will show in the next section,
using as an example the anharmonic oscillator, that it is possible to
take into account an important part of the correlations present in the
exact ground state in a fully consistent way.

\section{The model in standard RPA}
\label{RPAnormal}
We start from the Hamiltonian of an anharmonic oscillator with quartic
coupling:
\beq 
\mathrm{H} = \frac{p^2}{2} + \mu^2 \frac{X^2}{2} + g X^4~.
\eeq
with constants $\mu^2 <0$ and $g> 0$.
In the following sections, $\tilde{\mu} \equiv \sqrt{|\mu^2|}$.
This Hamiltonian can be rewritten as
\beq 
\mathrm{H} =  \frac{p^2}{2} + \frac{X^2}{2} + (\mu^2 - 1) \frac{X^2}{2} + g X^4
\eeq
and can be seen as a harmonic oscillator disturbed by a potential 
$V(X) = (\mu^2 - 1) \frac{X^2}{2} + g X^4$.
We introduce the usual destruction and creation
operators $a$ and $a^\dagger$ defined by $a = (\tilde{\mu} X+ip)/
\sqrt{2\tilde{\mu}}$. 
The basis for standard RPA is the mean-field approximation which is
usually formulated with the help of a Bogoliubov
transformation. Thereby the starting point is the following trial wave
function (see, e.g., Ref.~\cite{BF88}):
\beq 
|\mathrm{HF}\rangle  = e^{(u a^{\dagger}+\frac{1}{2}ta^{\dagger 2})} |V\rangle ~,
\label{toto}
\eeq
with variational parameters $u$ and $t$. $|V\rangle $ is the vacuum of the
initial destruction operator $a$: $a|V\rangle  =0$. The destruction operator
of the Hartree-Fock vacuum $b |\mathrm{HF}\rangle = 0$ can be expressed with
the help of $u$ and $t$:
\bea
b=(1-t^2)^{-1/2}(a-t a^{\dagger} -u)~.
\eea
For later convenience we define $\omega\equiv \tilde{\mu} (1-t)/(1+t)$ and
$s\equiv \sqrt{2/\tilde{\mu}}\,u/(t-1)$. In terms of the new creation and
destruction operators we then obtain 
\bea 
x \equiv X+s &=& \frac{1}{\sqrt{2\omega}}(b+b^{\dagger})~, \nonumber \\ 
p &=& -i\sqrt{\frac{\omega}{2}}(b-b^{\dagger})~,
\label{xpdefine}
\eea
which makes the role of the variational parameters obvious: $s$ simply
describes a translation in space and $\omega$ a modified frequency of
the oscillator. The Hamiltonian can be cast into the
following form:
\bea 
\mathrm{H} = -\frac{\omega}{4}(b-b^{\dagger})^2 + \frac{ \mu^2+12 g
s^2}{4 \omega} (b + b^\dagger)^2 \nonumber \\
        - \frac{s \mu^2 +  4 g s^3 }{\sqrt{2 \omega}}(b+b^{\dagger})
- \frac{4 s g}{\sqrt{8 \omega^3}} (b+b^{\dagger})^3 + 
  \frac{g}{4 \omega^2} (b+b^{\dagger})^4 \nonumber \\
  + g s^4 + \frac{\mu^2 s^2}{2}~.
\label{hhub}
\eea
The aim of the Bogoliubov transformation is to select the best basis
in the sense that the parameters $\omega$ and $s$ minimize the mean-field
ground state energy, $E^{HF}_0
= \langle \mathrm{HF}| \mathrm{H}|\mathrm{HF}\rangle$. These two
parameters are thus determined by: $\frac{\partial E^{HF}_0}{\partial s} =
0, \frac{\partial E^{HF}_0}{\partial \omega} = 0$.  As can easily be
checked by a direct calculation, the above equations are equivalent to
the so-called ``gap equations'':
\bea 
\frac{\partial E^{HF}_0}{\partial s} = 0 &\Leftrightarrow& \langle \mathrm{HF}|
[\mathrm{H},b]|\mathrm{HF} \rangle = 0~, \nonumber \\ 
\frac{\partial E^{HF}_0}{\partial \omega} = 0 &\Leftrightarrow& \langle \mathrm{HF}|
[\mathrm{H},b\,b]|\mathrm{HF} \rangle = 0~.
\label{gapmf}
\eea
If $|\mathrm{HF}\rangle $ was the exact ground state, the above gap
equations would be satisfied automatically. However, the deeper origin
of these equations (minimization of the ground state energy with respect to the
Bogoliubov parameters $\omega$ and $s$) implies that they have always to be
satisfied.

Let us now discuss the RPA corrections to the ground state energy.  In
standard RPA the excitation operator $A$ entering the RPA equations,
Eqs.~(\ref{RPAEQC}), is taken to be $A \in \{b,b^\dagger,bb,b^\dagger
b^\dagger\}$.  We define the RPA ground state energy as follows:
\beq
E_0^\mathrm{RPA} = \langle \mathrm{RPA}|\mathrm{H}|\mathrm{RPA} \rangle~.
\eeq
This expression involves expectation values of the type 
$\langle \mathrm{RPA}|A \,B|\mathrm{RPA}\rangle$.
These RPA corrected values can 
be calculated from the RPA Green's functions in the limit $t \rightarrow t'$:
\beq
\langle \mathrm{RPA}|A \,B|\mathrm{RPA}\rangle = i \int \frac{dE}{2\pi} G_{AB}(E)~.
\label{sumrpa}
\eeq
It is in this way that we introduce RPA fluctuations in the calculation of 
the ground state energy.

In standard RPA the Green's functions on the right hand side of
Eq.~(\ref{sumrpa}) are not determined self-consistently, i.e., the expectation
values entering the Green's functions themselves are determined using
the mean-field ground state $|\mathrm{HF}\rangle$.
In our simple toy model we can obtain analytic expressions for
the Green's functions.

Note that from Eqs.~(\ref{RPAEQC}) we do not directly obtain Green's functions
like $G_{x,x^2}$ or $G_{p,p}$, for example, since the operators $x$, $x^2$,
and $p$ are not contained in our set of excitation operators $\{
b,b^\dagger,bb,b^\dagger b^\dagger\}$. Hence, in order
to apply Eq.~(\ref{sumrpa}) to expectation values like $\xxx$ or
$\pp$, the operators $x$ and $p$ are expressed in terms of $b$
and $b^\dagger$
(cf. Eq.~(\ref{xpdefine})), and the resulting expectation values like
$\langle b b b\rangle$, $\langle b^\dagger b b\rangle$, $ \langle b
b\rangle$, etc., are obtained from the Green's functions $G_{b,b b}$,
$G_{b^\dagger,b b}$, $G_{b,b}$, etc.\footnote{Note that there is an ambiguity
in the assignment of the time arguments to some of the Green's functions.
This does, however, not influence the results (see Appendix~\ref{app} for a
more detailed discussion).} The short-hand notation $G_{x^3}$ or
$G_{p^2}$ has to be understood as the corresponding linear combination of
Green's functions like $G_{b,b b}$, $G_{b^\dagger, b b}$, $G_{b,b}$, etc.,
needed for the calculation of $\xxx$ or $\pp$.

To determine $E_0^\mathrm{RPA}$ we need, in addition to $\xx$,
$\pp$, and $\xxx$, also the expectation value $\xxxx$. However, 
$G_{x^4}$ contains Green's functions of the type
$G_{b^{\dagger},bbb}$, that is, $A \in \{
b,b^\dagger,bb,b^\dagger b^\dagger\}$ is not sufficient to calculate
$\xxxx$ \footnote{Since the excitation operators have to be
non-hermitian, two-particle operators of the form $b^\dagger b$ are
not allowed. Thus we cannot use, e.g., $G_{b^\dagger b,bb}$ to
determine $G_{x^4}$.}.  We thus approximate this term by a
factorization:
\beq
\langle \mathrm{RPA}|x^4|\mathrm{RPA}\rangle \simeq 
                                 3 {\langle
\mathrm{RPA}|x^2|\mathrm{RPA}\rangle}^2~. 
\eeq
Finally, note that in the symmetric case ($s=0$), the three-operator
expectation values vanish which implies that the Green's function are
the Hartree-Fock ones:
\beq
i \int \frac{dE}{2\pi} G^{s=0}_{x^2}(E) = \langle \mathrm{HF}|x^2|\mathrm{HF}\rangle
\eeq
Hence in this case the RPA energy is equal to the mean-field one.

In the literature the term RPA is used for a similar approximation to
the Green's functions as described above. But, it has to be mentioned
that often the charging formula~\cite{HCDS02,FW71},
\beq
E_0 = E_\mathrm{HF} + \int_0^1 \frac{d\lambda}{\lambda} \langle\lambda|
\mathrm{H}_\mathit{int}(\lambda)|\lambda\rangle~,
\label{charging}
\eeq
is applied to determine the ground state energy. 
Thereby $\mathrm{H}_\mathit{int}(\lambda) = \lambda
\mathrm{H}_\mathit{int}$ and the state $|\lambda\rangle$ is defined as
the ground state with respect to $\mathrm{H}(\lambda) =
\mathrm{H}_\mathit{bare} + \lambda \mathrm{H}_\mathit{int}$ where
$\mathrm{H}_\mathit{int}$ is the interacting part of the hamiltonian.

But, what we want to do is to compute directly $\langle gs |\mathrm{H} | gs
\rangle$ with $| gs\rangle = | \mathrm{HF}\rangle, |\mathrm{RPA}\rangle$ or
$| 0\rangle$ (see next section) and compare the result obtained within the
different approximations. Thus again, $E_{0}^{\mathrm{RPA}}$ is not the energy
as understood in standard RPA but directly $\langle \mathrm{RPA}|\mathrm{H}
|\mathrm{RPA}\rangle$.
 
\section{The renormalized RPA approach}
\label{RRPA}
What we want to do now is to go beyond the standard RPA approach and
retain some of the correlations in the ground state. The starting
point will be again the Bogoliubov transformation described in
Sec.~\ref{RPAnormal}. In contrast to standard RPA we will, however,
not use the mean-field ground state $|\mathrm{HF}\rangle$ but a more complicated
one denoted $|0\rangle$. This new ground state, which will not be constructed
explicitly in the following, contains some correlations. Since we
want to use the ``best'' one, the parameters $\omega$ and $s$ will be
again determined by the minimization of the ground state energy $E_0
= \langle 0|\mathrm{H}|0\rangle $. This leads to generalized gap equations:
\bea 
\frac{\partial E_0}{\partial s} = 0 &\Leftrightarrow& \langle 0|
[\mathrm{H},b]|0 \rangle = 0~, \label{gaps} \\ 
\frac{\partial E_0}{\partial \omega} = 0 &\Leftrightarrow& \langle 0|
[\mathrm{H},b\,b]|0 \rangle = 0~.
\label{gapw}
\eea 
which are analogous to $\langle 0|[\mathrm{H},Q^{\dagger}_{\nu}]|0
\rangle =0$ with the formalism introduced in Sec.~\ref{RPAeqs}. We may
note that the equivalence between these formulas is no longer valid if
we consider three particle excitation operators (see Sec.~\ref{modifRRPA} for
further details).

More precisely, independent of the approximation made to include
correlations in the ground state, the above equations will guarantee that
the basis is the energetically most favored one whithin this approximation.

In the remaining part of this article we will assume that all
expectation values of the type $\langle 0|bb|0\rangle$, $\langle
0|bb^{\dagger}|0\rangle$, $\langle 0|bbb^{\dagger}|0\rangle $,
etc. are real. This implies for example that $\langle 0|p\,x|0\rangle = -i/2$
or $\langle 0|p\,x^2|0\rangle = 0$,
independently of the specific ground state used. Thus, the key
point is to determine the remaining expectation values in the a priori
unknown ground state $|0\rangle$. In renormalized
RPA~\cite{R68,H64,CDS94,TS95} this is achieved with the use of Green's
functions in the limit $t \rightarrow t'$. More precisely, we get
self-consistency conditions for expectation values with two operators, i.e.,
$\langle0|x\,x|0\rangle$ and $\langle0|p\,p|0\rangle$ from
\beq
\langle 0| A \,B|0\rangle = i \int \frac{dE}{2\pi} G_{AB}(E)~.
\label{sum}
\eeq
In contrast to Eq.~(\ref{sumrpa}), the expectation values entering the
Green's functions on the right hand side of the above equation are
determined self-consistently, i.e., not within the Hartree-Fock ground state.
The remaining expectation values are approximated using a factorization, e.g.,
\beq
\langle 0|p^2 x^2|0\rangle =
  \langle 0|p^2|0\rangle \langle 0|x^2|0\rangle +
  2 \langle 0|px|0\rangle^2~.
\eeq
This means that the self-consistent procedure is limited to expectation values
with at most two operators and that the correlated part of the expectation
values with more than two operators is neglected~\cite{HCDS02}. Moreover, since
$\langle x\rangle $ vanishes, it follows immediately that terms with three
operators such as, e.g., $\xxx$ vanish.

As for the standard RPA analytic expressions for the Green's functions
can be given in terms of $\xx, \pp$ and $\px$ (see
Appendix~\ref{app}). The latter expectation values have then, in
principle, to be determined numerically via the self-consistency
conditions, Eq.~(\ref{sum}). With these values at hand, we are able,
in principle, to calculate the ground state energy $E_0$. However,
renormalized RPA is known to miss an important part of the correlation
energy~\cite{HMDS02}. Besides, this method cannot be derived via a
variational principle and therefore the Ritz criterion is no longer
applicable. In fact, what happens in most cases, is that the ground
state obtains overbinding~\cite{HMDS02}. In our case it is even worse
since one can show that a contradiction with the virial theorem and
the Heisenberg uncertainty principle occurs. To that end let us look
at the explicit equation for $\pp$, which is obtained by integrating
$G_{p^2}(E)$, given in Eq.~(\ref{gppx}), over $E$.
The result is:
\bea 
\pp = i \int \frac{dE}{2 \pi} G_{p^2}(E) = \epsilon^2 \xx \nonumber \\ + i
\frac{4 ( 12 s g)^2 \pp}{\epsilon^2} \int \frac{dE}{2 \pi} \frac{1} \nonumber
{ (E^2 - E_1^2) (E^2-E_2^2)}~,
\eea
where $\epsilon, E_1$ and $E_2$ are defined in Eq.~(\ref{abbrev}).  If
we now look at the explicit form (Eq.~(\ref{gapwa})) of the gap
equation for $\omega$ we conclude that the second term on the right
hand side should vanish. However, this is only possible for $\pp$ = 0,
implying a vanishing kinetic energy. This is obviously not a
physically reasonable solution as it is in contradiction to the virial
theorem. Even if we forget for the moment this contradiction, the
self-consistent procedure can not give the right values for the
occupation numbers and consequently for the energy. A possible way to
circumvent that problem, that is to compute correctly the energy, is
to use the ``charging formula'' as developed, e.g., in \cite{HCDS02}.
It is applied in a similar way as in standard RPA
(cf. Sec.~\ref{RPAnormal}) to compute the correlation energy. The main
difference is that we now have in principle a closed system of
equations which have to be solved self-consistently. Following
\cite{HCDS02}, one therefore replaces the gap equation concerning $s$,
Eq.~(\ref{gaps}), by a minimization of the ground state energy
obtained with the help of the charging formula
(cf. Eq.~(\ref{charging})).  This has of course one disadvantage: One
loses the symmetry of the double commutators, i.e.
\beq 
\langle [b,[\mathrm{H},b^\dagger
b^\dagger]]\rangle - \langle [b^\dagger
b^\dagger,[\mathrm{H},b]]\rangle = \langle [\mathrm{H},b^\dagger]
\rangle \neq 0~.  
\eeq
This symmetry property, however, is crucial for the stability of the
system~\cite{TS02}. We will therefore proceed in another direction.

\section{The extended r-RPA approach}
\label{modifRRPA}
Instead of using the charging formula, our strategy to include further
correlations in a minimal way will be as follows: We will abandon the
factorization approximation for the lowest non-vanishing expectation
values of products of more than two operators. 

For the deformed case ($s\neq 0$) this means that we will keep the
values of $\xxx$ and $\ppx$ (these are the only non-vanishing
expectation values of three operators).  Similar to $\xx$ and $\pp$,
their actual values are determined via the self-consistency
conditions, Eq.~(\ref{sum}). With these expectation values at hand
the ground state energy is then obtained, as before, as the
expectation value of the Hamilton operator (cf. Eq. (\ref{hhub})): 
\beq 
E_0^{s\neq0} =
\frac{\pp}{2} + \frac{\mu^2}{2}(\xx + s^2)+ g (3 \xx^2+s^4+ 6 s^2
\xx-4 s \xxx)~.  
\eeq 
First of all we have to note that there is no longer any obvious
contradiction with the virial theorem. Moreover, as we will see in the
next section, as far as we are not too close to the ``phase
transition'', our results for the ground state energy are very close to
the exact ones, without any further manipulation, like, e.g., the
charging formula. This means $\xxx$ and $\ppx$ contain already an
important part of the so far missing correlations.  This does not
necessarily imply that we also reproduce the exact values for, e.g.,
$\xx$ or $\pp$. But the correlations are combined in such a way that
the energy is determined very precisely within this approximation.

If we try to go further and include, e.g., also the correlated part of
expectation values with four operators, we face serious problems. To
obtain the Green's functions necessary to calculate $\xx,
\pp,\px,\xxx$ and $\ppx$ it suffices to restrict the space of
excitation operators to $A = \{b,b^\dagger,bb,b^\dagger b^\dagger\}$,
i.e., two-particle excitations. As long as we stay in that space, the
gap equations together with the assumption of real expectation values
for $\langle bb^\dagger\rangle$, etc., guarantees the symmetry of the
double commutators. As already discussed in Sec.~\ref{RPAnormal}, we
need Green's functions of the type $G_{b^{\dagger},bbb}$ to obtain,
e.g., $\xxxx$. That is, three particle excitation operators have to
included. In addition to the (practical) computational problems we
encounter another difficulty: As mentioned in Sec.~\ref{RPAnormal},
the Bogoliubov transformation contains only two parameters and
consequently, including three particles implies a priori an
approximation for the double commutators since there is no relation to
enforce their symmetry. For instance, to ensure the following symmetry
relation
\beq 
\langle [b b ,[\mathrm{H},b ^\dagger b^\dagger
b^\dagger]]\rangle - \langle [b^\dagger b^\dagger
b^\dagger,[\mathrm{H},b b ]]\rangle = 6 \langle [\mathrm{H},b^\dagger
b^\dagger b ]
\rangle~,  
\eeq
we needed an equation which guarantees $\langle
[\mathrm{H},b^\dagger b^\dagger b ] \rangle = 0$. As mentioned above
the Bogoliubov transformation does not provide us with an additional
adjustable parameter. The
assumption of a generalized gap equation of the form $\langle
[\mathrm{H},Q]\rangle=0$ as used in many RPA approaches also did not help. 
In this context boson expansion
techniques~\cite{ASW97,A00}, mapping a pair of bosons onto one new
boson, could be helpful. 

In the symmetric case $s=0$, it is known that, by parity, $\xxx$ and $\ppx$
vanish. Following the same underlying philosophy as in the deformed
case, we are now led to include $\xxxx, \pppp$ and $\ppxx$.  This is, as
stated above, a subtle task since the double commutators need no
longer to be symmetric.  Fortunately, this difficulty can be overcome:
Although in principle there are inherent problems with the symmetry of
the double commutators, one can explicitly check that all
non-symmetric terms disappear for $s=0$. This enables us to solve the
Dyson equation without symmetrizing by hand (as it is often done in
nuclear physics \cite{TS02}). The corresponding Green's functions are
listed in Appendix~\ref{app}. The expectation values $\xx,
\pp,\xxxx,\pppp$ and $\ppxx$ are then determined self-consistently
using Eq.~(\ref{sum}). In this case the ground state energy is given
by:
\beq
E_0^{s=0} = \frac{\pp}{2} + \frac{\mu^2}{2} \xx + g \xxxx~.
\eeq
As can be seen in the next section, the
numerical results for the ground state energy are in very good
agreement with the exact ones.

We have to emphasize that our approach is not fully self-consistent in
the sense of SCRPA, although the expectation values of up to three or
four operators, respectively, are calculated self-consistently. Let us
mention one point 
which enlightens the difference of our method to the SCRPA
approach used in Ref.~\cite{DS91}. The authors use the following form
for the RPA excitation operator:
\beq
Q^\dagger = \frac{1}{\sqrt{2}} (\lambda b^\dagger b^\dagger - \mu b b)~.
\eeq 
This relation is easily inverted to give $b b$ and $b^\dagger
b^\dagger$ in terms of $Q^\dagger$ and $Q$. Then the relation 
\beq
Q |SCRPA\rangle = 0
\label{rpafund}
\eeq 
can be applied to show that, e.g.\@, $\langle b b\rangle$
vanishes. Our excitation operator contains in addition one and three
particle operators, proportional to $b, b^\dagger$ and $bbb, b^\dagger
b^\dagger b^\dagger$, respectively. This renders the
inversion less obvious and we therefore cannot directly make use of
Eq.~(\ref{rpafund}). Thus, a priori, we have to calculate several
expectation values which can be shown to vanish within the SCRPA
formalism of Ref.~\cite{DS91}.  
The SCRPA results for the ground state
energy in the symmetric case are also in very good
agreement with the exact ones~\cite{DS91}. 

\section{Numerical results}
\label{numerics}

\begin{figure}[t]
\includegraphics[width = 9.cm]{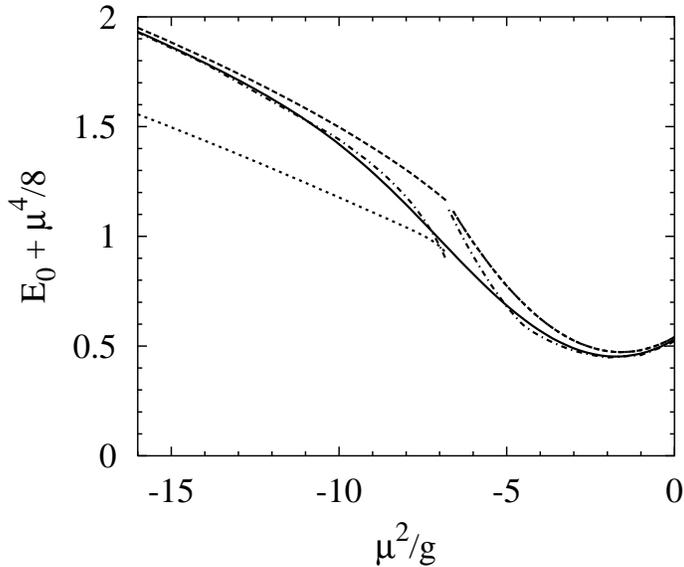}
\caption{Exact solution for the ground state energy (solid line), mean
field solution (dashed line), RPA solution (dotted line) and our
extended r-RPA solution (dashed-dotted line) as a function of
$\mu^2/g$ with $g=0.5$. Following \cite{BF88}, we added $\mu ^4 /8$ in order to
have positive energies. In the region where $s = 0$ the mean-field and
the RPA solution coincide (see text).}
\label{fige0}
\end{figure}

This section is devoted to a discussion of the numerical results.  In
Fig.~\ref{fige0} the energy of the ground state is displayed within
the different approaches (exact result, mean-field, RPA, and our
extended r-RPA result) as a function of $\mu^2 /g$. Of course, even the
mean-field is in reasonable agreement with the exact results (10\% at
worst). The largest discrepancy can be observed in the region near the
``phase transition'' at $\mu^2/g \approx -6.8$, i.e., where $s$ jumps
from some value $s \neq 0$ to $s = 0$. This is also the case for the RPA
and our r-RPA approach. But the figure 
clearly shows that the correlations incorporated in our approach are
important: The agreement with the
exact result is much better than in mean-field or RPA and
discrepancies show up only in a very small region. 
Note that, the curve obtained within our approach is not continuous at the transition
point. This is of course due to the difference in approximations made in the two
domains: in one case, $s\neq 0$, one has incorporated the effect of $\xxx$ and
$\ppx$, whereas in the other case, $s=0$, we have included
correlations in $\xxxx$, $\pppp$ and $\ppxx$. 
Near the transition, also higher order correlations become important,
and we therefore miss the exact result near the transition point. 

Before we come to the conclusion, we would like to give one explicit
example concerning the difference of our approach to the SCRPA: 
For $\mu^2/g = -1$, we obtain $\langle b^\dagger bbb \rangle = -0.033$
and $\langle b^\dagger b^\dagger bb \rangle = 0.026$, i.e., both
expectation values are of the same order of magnitude. In the SCRPA
approach used in Ref.~\cite{DS91} the former expectation value
vanishes identically.

\section{Conclusion}  
\label{conclusion}
We have considered
an extended version of renormalized RPA that yields 
values of high accuracy for the ground state energy of the anharmonic
oscillator. Moreover, the self-consistent procedure developed here
removes contradictions present in renormalized RPA, that is $\xx$
and $\pp$ can be determined without evidently violating the virial theorem or
the Heisenberg uncertainty principle. 

This means that the correlations included, $\xxx$ and $\ppx$ for $s
\neq 0$ and $\xxxx, \pppp$ and $\ppxx$ for $s =0$, represent the most
important part of the correlations missed in renormalized
RPA. Furthermore the systematics of the method allow us to think that
it can be easily applied to quantum field theory and cure some of the
problems which emerged in the work of Ref.~\cite{HCDS02}.  One problem
of our approach, on which we certainly should improve, is the fact
that we do not have a unified approach for the two cases $s=0$ and
$s\neq 0$.

Of course, it should be mentioned that there exist other
powerful methods such as CCM (Coupled Cluster Method) \cite{BF88} that
describe very accurately the properties of the anharmonic
oscillator. For problems concerning quantum chemistry or solid state
physics, this technically and numerically very involved method and its
extensions (see \cite{BF88} for further details) are probably the
most precise and well-suited. Actually, the approach developed in this
paper probably can not compete with CCM in all these
situations.  Nevertheless it has the great advantage to identify
clearly the underlying physics (correlations) and to be very easy to
use.

\section*{Acknowledgments:} 
We gratefully acknowledge stimulating discussions with Z. Aouissat,
and A. Rabhi. We also would like to thank G. Chanfray and P. Schuck
for their constant
interest in this work. One of us (M.O.) would like to thank the
Alexander von Humboldt foundation for financial support.

\begin{appendix}
\section{Explicit expressions for the Green's functions}   
\label{app}
We present here the analytical results for the various Green's functions needed
throughout the calculation. 
Note that the Green's function listed
below do not directly emerge from the solution of the system of
Eqs.~(\ref{RPAEQC}) but we have to reexpress 
operators containing $x$ and $p$ with $b$ and
$b^\dagger$, cf. Eqs.~(\ref{xpdefine}). For instance
\bea
G_{:x^4:}(E) &=& \frac{1}{4 \omega^2} G_{:(b +
b^\dagger)^4:}(E)\nonumber \\ &=&
\frac{1}{4 \omega^2}( G_{b b,b b}(E) + G_{b^\dagger
b^\dagger,b^\dagger b^\dagger}(E) + 2 G_{b^\dagger b^\dagger, b b}(E)
\nonumber \\ & & \hspace{-.5cm}+ 4 G_{b^\dagger, b b b}(E) +
4 G_{b^\dagger b^\dagger b^\dagger,b}(E)
+ 4 G_{b^\dagger b^\dagger, b b}(E)) ~.
\eea
One point is worth noting: We are dealing (in configuration
space) with two-time Green's functions and the notation $G_{A,B}(E)$
indicates that the time arguments for the corresponding Fourier
transformed quantity $G_{A,B}(t-t')$ are assigned as $A(t), B(t')$.
This means that there is an ambiguity in the assignment of time arguments
to the Green's functions which contain more than two operators. In
most cases this can be overcome by the fact that our excitation
operators are limited to non-hermitian ones, i.e.\@ an operator like 
$b^\dagger(t) b(t)$ cannot exist. But, 
$G_{b b,b b}(E)$ as well as $G_{b, b b b}(E)$ do exist and in general they
do not lead to the same expression. We are, however, only
interested in the limit $t \rightarrow t'$ to determine the
expectation values at equal times. In this limit all functions coincide.
\subsection{Symmetric case}
In the symmetric case, i.e., $ s=0$, we obtain 
\bea
G_{x^2}(E) &=& \frac{(E^2-\Sigma_{33})-9 \frac{(\eta-\nu)\sigma^2}{4
\alpha_3^2 \omega} }{(E^2-E_1^2) (E^2-E_2^2)}~, \nonumber \\
G_{p^2}(E) &=& \frac{\epsilon^2(E^2-\Sigma_{33})-9
\frac{(\eta+\nu)(\lambda-\omega^3\sigma)^2}{4 \alpha_3^2 \omega^5} }
{(E^2-E_1^2) (E^2-E_2^2)}~, \nonumber \\
G_{:x^4:}(E) &=& \frac{1}{4 \omega^2}\left(-\frac{2 (\lambda-2 g \omega+ 3
\omega^3(\alpha_2 E +\Omega))}{w^3 (E^2-E_{01} E_{02})} \right. \nonumber \\
&& \left. +
\frac{\tau}{4 \alpha_3^2 \omega^6 
(E^2-E_1^2) (E^2-E_2^2)}\right)~,\nonumber \\
G_{:p^4:}(E) &=& \frac{\omega^2}{4}\left(-\frac{2 (\lambda-2 g \omega+ 3
\omega^3(\alpha_2 E +\Omega))}{w^3 (E^2-E_{01} E_{02})} \right. \nonumber \\
&& \left. -
\frac{\tau}{4 \alpha_3^2 \omega^6 
(E^2-E_1^2) (E^2-E_2^2)}\right)~,\nonumber \\
G_{:p^2x^2:} (E) &=& \frac{\alpha_2}{2 \omega^3} \frac{E +
E_{02}}{E^2-E_{01} E_{02}}~, 
\label{greensym}
\eea
where we have defined the following abbreviations:
\bea
\epsilon^2 &=& \mu^2 + 12 g \xx~, \nonumber \\
\alpha_2 &=& \frac{\pp + \omega^2 \xx}{2 \omega} ~, \nonumber\\
\alpha_3 &=&\frac{9}{2} \nppxx-3 + \frac{9}{ 4 \omega^2}\left( \npppp + 4
\pp \omega \right. \nonumber \\ & & \left. + 4 \omega^3 \xx + \omega^4
\nxxxx\right )~,\nonumber\\
\sigma &=& \pp - \omega^2 \xx~,\nonumber \\
\lambda &=&  2 g \Big(-3 \pp(1-2 \omega \xx)+ \omega(-2 \nonumber \\&& 
-6 \nppxx+3
\omega (5 \xx-2 \omega \xx^2\nonumber \\ && +2 \omega \nxxxx))\Big) + \omega
(\omega^2-\epsilon^2) \sigma~,\nonumber \\
\Sigma_{33} &=& \frac{\eta^2-\nu^2}{\alpha_3^2}~,\nonumber \\
\eta &=& \langle\left[bbb,[\mathrm{H},b^\dagger b^\dagger b^\dagger]\right]
\rangle~,\nonumber\\
\nu &=& \langle \left[bbb,[\mathrm{H},b b b]\right]\rangle~,\nonumber\\
\Omega &=& \frac{1}{\omega^3}\Big(\lambda-4 \omega^3(\pp+\epsilon^2
\xx)+22 g \omega\nonumber \\ &&+16 \omega^2 g (-6 \xx+3 \omega \xx^2-2 \omega
\nxxxx)\Big) ~,\nonumber \\
\tau &=& 6 \alpha_3\Big(9 \lambda^2 \sigma + \lambda\left(-27
\sigma^2 + 4 (\alpha_3 E + \eta + \nu) (E-\omega)\right)
\omega^3\nonumber\\ && + 2
\sigma \omega^5\left(\omega\left(-4 E (\alpha_3 E + \eta) + 9 \sigma^2
\right. \right. \nonumber\\ && \left. \left.
+ 2 (\alpha_3 E + \eta +\nu) \omega \right) + 2 (\alpha_3 E + \eta -
\nu) \epsilon^2\right)\Big)~,\nonumber \\
E_{01,2} &=& \frac{\Omega}{\alpha_2} \pm \frac{\lambda-2 g
\omega}{\alpha_2 \omega^3}~.
\eea
The energies $E_1^2$ and $E_2^2$ are the solution of the following
equation:
\bea
E^4 + \frac{E^2}{4 \alpha_3 \omega^3}\left(18 \sigma (\lambda-\sigma
\omega^3)- 4 \alpha_3 \omega^3 (\Sigma_{33}+\epsilon^2)\right)\hspace{5cm}&&
\nonumber \\
+ \frac{1}{16 \alpha_3^2 \omega^6} \left( (81 \sigma^2+36 \omega (\eta
+ \nu) ) (\lambda - \sigma \omega^3)^2 \right. \hspace{5cm} && \nonumber \\
\left. 
+ \epsilon^2 \omega^5(36
\sigma^2(\eta-\nu)+16 \alpha_3^2 \omega \Sigma_{33})\right) = 0~.\hspace{5cm}&&
\eea 
$\xxx$ and $\ppx$ vanish because
of parity. $\langle : \mathcal{O}:\rangle$ denotes the expectation value
of the normal ordered product of the operator $\mathcal{O}$. If
$|\mathrm{HF}\rangle$ is taken as ground state those expectation
values obviously vanish and $\xx = 1/(2 \omega), \pp = \omega/2$. From
the gap equations, Eqs.~(\ref{gaps},\ref{gapw}), we obtain the following condition
\beq 
\omega^2 (\pp - \epsilon^2 \xx) + 3 g-12 g \omega \xx(1-\omega \xx)- 4 g \omega^2 \nxxxx = 0~.
\eeq
\subsection{Deformed case}
In the deformed case, i.e.\@ $s \neq 0$, 
we obtain:
\bea
G_{x^2}(E) &=& \frac{E^2 - \frac{\beta \kappa}{\alpha_2^2}}{ (E^2 -
E_1^2) (E^2-E_2^2)} \nonumber \\ & = & \frac{1}{E^2-\epsilon^2+\frac{9 g^2
\delta^2 (\beta-\kappa)}{\omega^2 \alpha_2^2}
\frac{1}{E^2-\frac{\beta\kappa}{16 \omega^4 \alpha_2^2}}}~, \nonumber \\
G_{p^2}(E) &=& \epsilon^2 G_{x^2}(E) + \frac{288 \delta^2 g^2 \kappa}
{\alpha_2^2} \frac{1}{ (E^2 - E_1^2) (E^2-E_2^2)}~, \nonumber \\
G_{:x^3:}(E) &=& -\frac{6 \delta g\left(\alpha_2^2 E^2 + 4 \omega \kappa
+ 3 E ( \kappa+ \alpha_2 \omega)\right)
}{\alpha_2\omega^2 (E^2 - E_1^2) (E^2-E_2^2)}~, \nonumber \\ 
G_{:p^2x:}(E) &=& \frac{6\delta g \left( \alpha_2^2 E^2- E (\kappa + 
\alpha_2 \omega)\right)}{\alpha_2 (E^2 - E_1^2) (E^2-E_2^2)} ~,
\label{gppx}
\eea
where we have introduced in addition the following abbreviations:
\begin{eqnarray}
\epsilon^2 &=& \mu^2 + 12 g (s^2 + \xx) \nonumber \\
\beta &=& 4 \left( \pp + \xx (\epsilon^2 + 12 g \xx) - 18 g s \nxxx\right)\nonumber \\
\delta &=& 2 s \xx - \nxxx \nonumber \\
\kappa &=& 2 \left(\pp ( 1 + \frac{\epsilon^2}{\omega^2}) - 12 g s \nxxx
+(\omega^2+\epsilon^2) \xx \right. \nonumber \\ & & \left.
- \frac{24 g}{\omega^2} \nppx s\right) \nonumber \\
E_{1,2}^2 &=& \frac{1}{2} \left( \epsilon^2 + \frac{\beta \kappa}{\alpha_2^2}
\pm \sqrt{ (\epsilon^2 - \frac{\beta \kappa}{\alpha_2^2})^2
+ 8 ( 12 g \delta)^2 \frac{\kappa}{\alpha_2^2}} \right)~.
\label{abbrev}
\end{eqnarray}  
In this case we end up with two gap equations, one determining the
value of $s$,
\beq
s^3 + s \left( 3 \xx + \frac{\mu^2}{4 g}\right) - \nxxx = 0~,
\label{gapsa}
\eeq
and another one concerning $\omega$,
\beq
\pp = \epsilon^2 \xx - 12 g s \nxxx~.
\label{gapwa}
\eeq
Note that the expression for $G_{x^2}(E)$ given in Eq.~(\ref{gppx}) can
be interpreted as a free ``propagator'' with an additional
``self-energy'' depending on $E$. Similar interpretations are possible
for all the other expressions.
\end{appendix}

\end{document}